\documentclass[11pt]{article}

\usepackage{amsmath,amssymb,amsfonts,amsthm}
\usepackage{hhline}
\usepackage{tikz}
\usepackage{subfigure}
\usepackage[margin=1in]{geometry}

\newtheorem{theorem}{Theorem}
\newtheorem{definition}[theorem]{Definition}
\newtheorem{lemma}[theorem]{Lemma}
\newtheorem{proposition}[theorem]{Proposition}
\newtheorem{corollary}[theorem]{Corollary}
\newtheorem*{theorem*}{Theorem}
\newtheorem*{lemma*}{Lemma}

\theoremstyle{remark}
\newtheorem*{remark*}{Remark}
\newtheorem{observation}[theorem]{Observation}

\usepackage{geometry}
\usepackage[noline, ruled,linesnumbered,commentsnumbered]{algorithm2e}
\usepackage{todonotes}

\newcommand{\abs}[1]{\left\vert#1\right\vert}
\newcommand{\set}[1]{\left\{#1\right\}}
\newcommand{\tuple}[1]{\left(#1\right)}

\newcommand{\defeq}{\triangleq}

\newcommand{\wt}[1]{\mathrm{wt}\tuple{#1}}

\renewcommand{\mid}{\;\middle\vert\;}

\usepackage{xifthen}

\renewcommand{\Pr}[2][]{ \ifthenelse{\isempty{#1}}
  {\mathbf{Pr}\left[#2\right]} {\mathbf{Pr}_{#1}\left[#2\right]} }
\newcommand{\E}[2][]{ \ifthenelse{\isempty{#1}}
  {\mathop{\mathbf{E}}\left[#2\right]}
  {\mathop{\mathbf{E}}_{#1}\left[#2\right]} }

\usepackage{hyperref}

\renewcommand{\exp}[1]{\mathrm{exp}\tuple{#1}}

\newcommand{\commentout}[1]{}
\newcommand{\eat}[1]{}
\newcommand{\even}{\mathsf{Even}}
\newcommand{\odd}{\mathsf{Odd}}

\newcommand{\calM}{\mathcal{M}}

\newcommand{\p}{\mathbf{p}}

\long\def\ignore#1{}

\begin{document}

\title{Canonical Paths for MCMC: from Art to Science}
\author{Lingxiao Huang  \thanks{Institute of Interdisciplinary Information Sciences, Tsinghua University. \url{huanglingxiao1990@126.com}. This author is supported in part by the National Basic Research Program of China Grant 2015CB358700, 2011CBA00300, 2011CBA00301, the National Natural Science Foundation of China Grant 61202009, 61033001, 61361136003.}\and Pinyan Lu \thanks{Microsoft Research. \url{pinyanl@microsoft.com}} \and Chihao Zhang \thanks{Shanghai Jiao Tong University. \url{chihao.zhang@gmail.com}. This author is supported in part by the National Natural Science Foundation of China Grant 61261130589, 61472239.} }
\date{}
\maketitle

\begin{abstract}
Markov Chain Monte Carlo (MCMC) method is a widely used algorithm design scheme with many applications. To make efficient use of this method, the key step is to prove that the Markov chain is rapid mixing. Canonical paths is one of the two main tools to prove rapid mixing. However, there are much fewer success examples comparing to coupling, the other main tool. The main reason is that there is no systematic approach or general recipe to design canonical paths. Building up on a previous exploration by McQuillan~\cite{mcquillan2013approximating}, we develop a general theory to design canonical paths for MCMC: We reduce the task of designing canonical paths to solving a set of linear equations, which can be automatically done even by a machine.

Making use of this general approach, we obtain fully polynomial-time randomized approximation schemes (FPRAS) for counting the number of $b$-matching with $b\leq 7$ and $b$-edge-cover with $b\leq 2$. They are natural generalizations of matchings and edge covers for graphs. No polynomial time approximation was previously known for these problems.
\end{abstract}

\section{Introduction}
In statistics and computer science, Markov Chain Monte Carlo (MCMC)
methods are a class of algorithms for sampling from a probability
distribution based on constructing a Markov chain that has the desired
distribution as its stationary (equilibrium) distribution. The state
of the chain after a number of (random) steps is then used as a sample
of the desired distribution. MCMC methods are primarily used for
calculating approximations of multi-dimensional integrals, number of
combinational objects, number of solutions for constraint satisfaction
problems, partition function for statistic physics systems and so
on~\cite{cryan2006rapidly,dyer1991random,app_DJV01,IS_DG00,app_GJ11,col_Jerrum95,jerrum1993polynomial,jerrum1996markov,app_JSV04,Morris02randomwalks,sinclair1992improved,col_Vigoda99}. Typically,
the support set of the distribution is exponentially large but we need
the sampling algorithm to run in polynomial time. This requires that
the Markov chain is rapidly mixing, namely, it is very close to the
stationary distribution after polynomial number of steps.

Canonical path is one of the two main tools (the other one is
coupling) to prove rapid mixing of the Markov chain. To make use of
this tool, one need to design paths between each pair of states for
the Markov chain and prove that the overall congestion at each link of
the Markov chain is low. However, it is typically a very difficult
task to come up with a low congestion routing especially for an
exponentially large state graph of a Markov chain. Thus, the design of
canonical paths for a given Markov chain remains a highly non-trivial
artwork for masters. For the other main tool coupling, there are quite
a few nice theories developed. One most important general approach is
path coupling~\cite{BD97} which enables one to only analysis the local
configuration of a single constraint rather than the global
configuration. This is typically much easier to handle.

Due to the lack of general theory and approach, there are only very
few notably successful examples of canonical path. One important
example is the MCMC for sampling and counting matchings of a
graph~\cite{jerrum1989approximating}. The states of the Markov chain
is all matchings for a given input graph.
The symmetric difference of two matchings of a graph is a disjoint
union of paths and cycles. Then, the natural and success canonical
path for matchings is ``winding" the edges one by one just follow the
natural order of these paths and cycles. Another important success
example is the so called ``sub-graph world" problem transformed from
ferromagnetic Ising model~\cite{jerrum1993polynomial}. For this
problem, the symmetric difference of two configurations can be any
graphs. But any graph has path-cycle decompositions, and their
canonical paths simply do an arbitrary path-cycle decomposition and
wind the edges following these paths and cycles. Since the constraint
in each vertex for that problem is the simple parity function, they
can prove that these canonical paths indeed have low congestion.

In an unpublished manuscript~\cite{mcquillan2013approximating},
McQuillan proposed a beautiful generalization of this path-cycle
decomposition idea called winding. In a high-level, one do not use a
single fixed path-cycle decomposition but use a convex combination of
exponentially many path-cycle decompositions and distribute the flow
among these canonical paths. This idea itself alone is not new, such
fractional canonical paths were used before, see for
example~\cite{Morris02randomwalks}. The main contribution
of~\cite{mcquillan2013approximating} is a method to design such a
convex combination by a local property for each constraint called
windable. As long as each local constraint is windable, they can
design the global path-cycle decompositions and thus canonical paths
automatically. Therefore, this winding approach gives a systematic
approach to design canonical paths for MCMC. This is similar to path
coupling technique for coupling which enables us to only analysis the
local constraint and configurations. However, to show that this
windable property for the local constraints still require a
construction for some mathematical objects. In their paper, they
showed that the Not-All-Equal functions satisfies the properties by an
explicit construction of these mathematical objects. It was not clear
how to show whether a new constraint function satisfies this windable
property or not.

In this paper, we give a characterization for the property of windable
by a set of linear equations, which works both for unweighed and
weighted constraints. Having that, the whole process of designing
canonical paths becomes a routine of solving linear equations which
can be automatically done by a machine.  We also refine some
definitions and presentation for the winding approach so that it is
easier to understand and apply. We extend this approach to instances
with edge weights as well.

It is very easy to verify that the matching
constraint~\cite{jerrum1989approximating} and parity
function~\cite{jerrum1993polynomial} are indeed windable by our
characterization. Moreover, with this powerful approach and
characterization in hand, we design a number of new fully
polynomial-time randomized approximation schemes (FPRAS) for
approximate counting by simply verifying that the local constraint
functions are windable by our new characterization theorem. Our first
example is counting $b$-matchings, which is a natural generalization
of matchings. A subset of edges for a graph is called a $b$-matching
if every vertex is incident to at most $b$ edges in the
set. $1$-matching is the conventional definition of matching for a
graph. In particular, we obtain FPRAS for counting $b$-matchings with
$b\leq 7$ for any graphs. Previously, FPRAS was only known for
counting $1$-matchings.

Another problem we resolve is a generalization of the edge cover
problem. A subset of edges for a graph is called an edge cover if
every vertex is incident to at least one edge in the set. Previously,
MCMC based approximation algorithm for counting edge covers was only
known for $3$-regular graphs~\cite{MFCS09}. In fact, they also used
canonical path to get rapid mixing and used path-cycle decomposition
to construct canonical paths. Since they do not have a systematic
approach but some ad-hoc construction and case-by-case analysis, they
only succeeded for the very special $3$-regular graphs. By our
approach and characterization, we can show that there exist a convex
combination of path-cycle decompositions which works for general
graphs. Moreover, we generalize it to $b$-edge-cover by requiring that
every vertex is incident to at least $b$ edges in the set. We obtain
FPRAS for counting $b$-edge-cover for $b\leq 2$. We note that FPTAS
based on correlation decay technique for counting edge covers for
general graphs was known~\cite{counting-edge-cover,LiuLZ14}. However,
it seems that their technique have intrinsic difficulty for
$2$-edge-cover.

Interestingly, we can show that the constraint function of
$8$-matchings and $3$-edge-cover are not windable by our
characterization theorem. We do not know whether these transitions
really corresponds to the boundaries of approximability or not. We
leave these as interesting open questions.

The most interesting future direction is to design canonical paths for
other Markov chains by this approach and thus get polynomial time
approximation algorithms. Of course, we are not claiming that winding
is the only way to design canonical paths. To develop other systematic
approach for designing and analyzing canonical paths for MCMC is very
interesting. We hope that our work can stimulate such kind research.


\section{Preliminaries}

\paragraph{Holant Problem.} Let $G(V,E)$ be a graph. In this
paper, we consider each edge $e=(u,v)\in E$ as two ``half edges''
$e_u$ and $e_v$\footnote{Here we consider ''half edges'' instead of 'edges' as usual, since our Markov chains work on these ''half edges''.}. Let $\mathcal{E}\defeq\set{e_u,e_v\mid e=(u,v)\in E}$
denote the set of all half edges. For every vertex $v\in V$, we use
$\mathcal{E}(v)$ to denote the set of half edges incident to $v$.

An instance of a Holant problem is a tuple
$\Lambda=\tuple{G(V,E),\tuple{f_v}_{v\in V}}$, where for every
$v\in V$, $f_v:\set{0,1}^{\mathcal{E}(v)}\to \mathbb{R}^+$ is a function, where $\mathbb{R}^+$ is the set of non-negative real numbers. For every assignment
$\sigma\in \set{0,1}^{\mathcal{E}}$, we define the weight of $\sigma$
as
\[
  w_{\Lambda}(\sigma)\defeq \prod_{v\in
    V}f_v\tuple{\sigma\mid_{\mathcal{E}(v)}}.
\]

For every $\sigma\in\set{0,1}^{\mathcal{E}}$, we use $d(\sigma)$ to
denote the number of edges $e=(u,v)$ such that $\sigma(e_u)$ and
$\sigma(e_v)$ disagree, i.e.,
$d(\sigma)\defeq\abs{\set{e=(u,v)\in E\mid \sigma(e_u)\ne
    \sigma(e_v)}}$.
For every $k\ge 0$, we denote
$\Omega_k\defeq\set{\sigma\in\set{0,1}^{\mathcal{E}}\mid d(\sigma)=k}$
and
$
  Z_k(\Lambda)\defeq\sum_{\sigma\in\Omega_k}w_\Lambda(\sigma).
$

The set $\Omega_0$ contains exactly all the assignments which are consistent at each edge.
These are the ordinary assignments we usually studied and we call  $Z(\Lambda)=Z_0(\Lambda)$ the \emph{partition function}
of $\Lambda$.

\paragraph{Symmetric Functions.}  A function
$f:\set{0,1}^J\to\mathbb{R}^+$ is \emph{symmetric}, if
the value of the function only depends on
the Hamming weight of its input.
We use $|x|=\sum_{i\in J}x_i$ to denote the Hamming weight of $x$.
Thus, for a symmetric function
$f:\set{0,1}^J\to \mathbb{R}^+$ where $\abs{J}=d$, we can write it as
$f=[f_0,f_1,\dots,f_d]$, where $f_i$ is the value of $f$ on inputs
with Hamming weight $i$.

We define some special symmetric functions which will be used in this
  paper:
  \begin{itemize}
  \item $\mathbf{0}$ ($\mathbf{1}$): $f(x)=0$ ($f(x)=1$) for all $x\in \{0,1\}^J$.
  \item $\even$ ($\odd$): $f(x)=1$ if $|x|$ is even (odd). Otherwise,
    $f(x)=0$.
  \item $= k$: $f(x)=1$ if $|x|=k$. Otherwise,
    $f(x)=0$.
  \item $\geq k$ ($\leq k$): $f(x)=1$ if
    $|x|\geq k$ ($|x|\leq k$). Otherwise,
    $f(x)=0$.
  \item $[a,b]$: $f(x)=1$ if $a\leq |x|\leq
    b$. Otherwise, $f(x)=0$.
  \end{itemize}
When needed, we use a sub index to indicate the arity of a function. For example, $\even_d$ and $(= k)_d$ is the $\even$ and $= k$ function with arity $d$. If every function $f_v$ is the function $(\leq 1)_{d_v}$, then the Holant problem $\Lambda=\tuple{G(V,E),\tuple{f_v}_{v\in V}}$ is the matching problem. Functions $\leq b$ are for $b$-matching problem and  functions $\geq b$ are for $b$-edge-cover problem.

We introduce a few operations for functions. For two functions $f$ and $g$ with same arity, we use $f\cdot g$ to denote the entry wise product of the two functions. For example:
   \begin{itemize}
  \item $[a,b]_d \cdot \even_d$:   $f(x)=1$ if $a\leq |x|\leq
    b$ and $|x|$ is even. Otherwise, $f(x)=0$.
      \end{itemize}
For a function $f:\set{0,1}^J\to\mathbb{R}^+$ and an
assignment $\pi\in \set{0,1}^I$ where $I\subseteq J$, we define the
\emph{pinning} of $f$ by $\pi$ as a function
$G:\set{0,1}^{J\setminus I}\to \mathbb{R}^+$ such that for every
$\sigma\in\set{0,1}^{J\setminus I}$, $G(\sigma)=f(\sigma\circ \pi)$
where $\sigma\circ\pi$ is the concatenation of $\sigma$ and $\pi$.
For symmetric functions in symmetric notation $[f_0,f_1,\dots,f_d]$,
a pinning gets a consecutive sub-sequence of $\{f_0,f_1,\dots,f_d\}$.
The complement of a function $\overline{F}$ takes a complement for each input entry before evaluation of the function.
For symmetric function, it simple reverses the order as $[f_d,f_{d-1},\dots,f_0]$.

\paragraph{Windable Functions.}
In \cite{mcquillan2013approximating}, a special family of functions
called \emph{windable functions} has been introduced:

\begin{definition}
  \label{def:wind}
  For any finite set $J$ and any configuration $x\in \set{0,1}^J$,
  define $\calM_x$ to be the set of partitions of $\set{i\mid x_i=1}$
  into pairs and at most one singleton.  A function
  $F: \{0,1\}^J\rightarrow \mathbb{R}^+ $ is \textbf{windable} if
  there exist values $B(x,y,M)\geq 0$ for all $x,y\in \{0,1\}^J$ and
  all $M\in \calM_{x\oplus y}$ satisfying:
  \begin{enumerate}
  \item $F(x)F(y)=\sum_{M\in \calM_{x\oplus y}}B(x,y,M)$ for all
    $x,y\in \{0,1\}^J$, and
  \item $B(x,y,M)=B(x\oplus S,y\oplus S,M)$ for all $x,y\in \{0,1\}^J$
    and all $S\in M\in \calM_{x\oplus y}$.
  \end{enumerate}
  Here $x\oplus S$ denotes the vector obtained by changing $x_i$ to
  $1-x_i$ for the one or two elements $i$ in $S$.  \footnote{Note that
    our definition seems different
    from~\cite{mcquillan2013approximating}, which defines $\calM_x$ to
    be the set of partitions of $\set{i\mid x_i=1}$ into pairs and
    singletons. While by the proof of Lemma 15
    in~\cite{mcquillan2013approximating}, both two definitions are
    equivalent to $F_{\oplus}$ being even-windable. Thus, our
    definition is equivalent to~\cite{mcquillan2013approximating} in
    fact.}
\end{definition}


\begin{observation}
  If $|x|$ is even, each $M\in \calM_x$ contains no
  singleton. Otherwise, if $|x|$ is odd, each $M\in \calM_x$ contains
  exactly one singleton.
\end{observation}

The following nice theorem was implicitly proved in
\cite{mcquillan2013approximating}.

\begin{theorem}
  \label{thm:rapidmixing}
  There exists an FPRAS to compute the partition function $Z(\Lambda)$
  for instances $\Lambda=\tuple{G(V,E),(f_v)_{v\in V}}$ with
  $\abs{V}=n$, if it holds that (1) the instance is
  \emph{self-reducible} in the sense of \cite{samp_JVV86}; (2) for every
  $v\in V$, the function $f_v$ is windable; and (3)
  $\frac{Z_2(\Lambda)}{Z_0(\Lambda)}=n^{O(1)}$.
\end{theorem}

The FPRAS is obtained by the MCMC method. The states of the Markov
chain are all the assignments in $\Omega_0 \cup \Omega_2$, which
contains all the consistent assignments ($\Omega_0$) and nearly
consistent assignments ($\Omega_2$). The second condition ensures that
the size of $\Omega_0$ and $\Omega_0 \cup \Omega_2$ are polynomial
related. To prove the rapid mixing of the Markov chain, the windable
condition is used to construct canonical paths. Roughly speaking, by
the pairings and singletons in the definition of windable, the graph
is naturally decomposed into disjoint union of paths and cycles. Then
the canonical path just winds the edges follow these paths and
cycles. The formal definition and detail can be found
in~\cite{mcquillan2013approximating}. For the convenience of the
readers, we also include a formal description for the Markov chain and
canonical paths in appendix. To logically follow the results of this
paper, all these are not needed except the statement of the above
theorem.

\section{Windability for Symmetric Functions}\label{sec:windable}
In this section, we obtain a characterization for all symmetric
windable functions. Before that, we introduce one more definition
which is also adapted from~\cite{mcquillan2013approximating}.

\begin{definition}
  \label{def:2decom}
  A function $H:\{0,1\}^J\rightarrow \mathbb{R}^+$ has a
  \textbf{2-decomposition} if there are values $D(x,M)\geq 0$, where
  $x$ ranges over $\{0,1\}^J$ and $M$ ranges over partitions of $J$
  into pairs and at most one singleton, such that:
  \begin{enumerate}
  \item $H(x)=\sum_{M}D(x,M)$ for all $x$, where the sum is over
    partitions of $J$ into pairs and at most one singleton, and
  \item $D(x,M)=D(x\oplus S,M)$ for all $x,M$ and all $S\in M$.
  \end{enumerate}
\end{definition}

Our definition for 2-decomposition is a generalization
of~\cite{mcquillan2013approximating}, since we allow the length of $J$
to be odd. By the new definition, we have the following lemma.

\begin{lemma}
  \label{lm:windto2}
  A function $F$ is windable, if and only if for all pinnings $G$ of
  $F$, the function $G\cdot \overline{G}$ has a 2-decomposition.
\end{lemma}

\begin{proof}
  If $F$ is windable, for each $I\subseteq J$ and each
  $\p\in \{0,1\}^I$, define
  $D_{\p}(x,M)=B((x,\p),(\overline{x},\p),M)$ for all
  $x\in \{0,1\}^{J\setminus I}$. By definition~\ref{def:wind}
  and~\ref{def:2decom}, we have that $D_{\p}$ is a 2-decomposition of
  $G\cdot \overline{G}$, where $G$ is the pinning of $F$ by $\p$.

  For the backwards direction, for all $x,y\in \{0,1\}^J$, let
  $I=\set{i\in J\mid x_i=y_i}$ be the position where $x$ and $y$
  agrees.  Let $\p\in \{0,1\}^I$ be the restriction of $x$ to $I$,
  which is the same as the restriction of $y$ to $I$.  Let $x'$ be the
  restriction of $x$ to $J\setminus I$. Define
  $B(x,y,M)=D_{\p}(x',M)$. Then by the definitions, it can be verified
  that $B$ witnesses that $F$ is windable.
\end{proof}


We introduce matrices $\mathbf{A}_m$ for every integer $m\ge 1$, which
will be used in our characterization theorem.

\begin{itemize}
\item If $m=2n$ is even, then
  $\mathbf{A}_m=(a_{ij})_{\substack{0\le i\le n\\0\le j\le n}}\in
  \mathcal{Q}^{(n+1)\times (n+1)}$ where
  \[
    a_{ij}=
    \begin{cases}
      \binom{i}{j}\binom{2n-i}{j}j!(i-j-1)!!(2n-i-j-1)!! &  \mbox{if $i\equiv j\pmod 2$;}\\
      0 & \mbox{otherwise}.
    \end{cases}
  \]
\item If $m=2n+1$ is odd, then
  $\mathbf{A}_m=(a_{ij})_{\substack{0\le i\le n\\0\le j\le n}}\in
  \mathcal{Q}^{(n+1)\times (n+1)}$ where
  \[
    a_{ij}=
    \begin{cases}
      \binom{i}{j}\binom{2n+1-i}{j}j!(i-j-1)!!(2n+1-i-j)!! & \mbox{if $i\equiv j\pmod 2$;}\\
      \binom{i}{j}\binom{2n+1-i}{j}j!(i-j)!!(2n-i-j) & \mbox{otherwise}.
    \end{cases}
  \]
\end{itemize}

The notation $n!!$ is the double factorial of $n$. For even $n$,
$n!!=n\cdot (n-2) \cdots 2$; and for odd $n$
$n!!=n\cdot (n-2) \cdots 1$.  If $n = 0$ or $n= -1$, then $n !! = 1$
by convention.  We note that $\mathbf{A}_m$ is a lower triangular
matrix (which follows from the convention that $\binom{i}{j}=0$ for
$i<j$).  The entry $a_{ij}$ of $\mathbf{A}_m$ has following
combinatorial interpretation: Consider we have $m$ balls consisting of
$i$ different red balls and $m-i$ different blue balls. If $m=2n$ is
even, then $a_{ij}$ is the number of ways to divide $2n$ balls into
$n$ pairs, such that the number of pairs with different colors is
$j$. If $m=2n+1$ is odd, then $a_{ij}$ is the number of ways to divide
$2n+1$ balls into $n$ pairs and a singleton, such that the number of
pairs with different colors is $j$.

\begin{lemma}\label{lem:2decom}
  Let $m\ge 1$ be an integer, $n=\lfloor\frac{m}{2}\rfloor$ and
  $H=[h_0,h_1,\dots,h_m]$ be a symmetric function with $h_i=h_{m-i}$
  for all $i=0,1,\cdots, n$. Let $\mathbf{h}=[h_0,h_1,\dots,h_n]$ be a
  vector.  Then $H$ is $2$-decomposible if and only if there exists an
  $\mathbf{x}\in\mathbb{R}^{n+1}\ge \mathbf{0}$ such that
  $\mathbf{A}_m\mathbf{x}=\mathbf{h}$.
\end{lemma}

We note that we abuse the notation $\mathbf{h}=[h_0,h_1,\dots,h_n]$
both as a symmetric function with arity $n$ and a vector in
$\mathbb{R}^{n+1}$ in the whole paper when meaning is clear from the
context.

\begin{proof}
  we first consider the case that $m=2n$ is even. Let $\mathcal{M}$
  denote the set of all partitions of $[m]$ into pairs.  We define an
  equivalent relation $\sim$ between pairs $(x,M)$ where
  $x\in\set{0,1}^m$ and $M\in\mathcal{M}$. Given a pair $(x,M)$, let
  $k(x,M)\defeq \abs{\set{(x_i,x_j)\in M\mid x_i\ne x_j}}$, i.e., the
  number of pairs in $M$ with different value. Then two pairs
  $(x,M)\sim (x',M')$ if $k(x,M)=k(x',M')$, namely $M$ and $M'$
  contain the same number of pairs with different value.  This
  relation induces equivalent classes $\set{\Delta_k\mid k=0,\dots,n}$
  where each $\Delta_k=\set{(x,M)\mid k(x,M)=k}$.

  We claim that the function $H$ is $2$-decomposible if and only if
  for every $0\le k\le n$, there exists $D_k\ge 0$ such that for every
  $x\in \set{0,1}^m$, $H(x)=\sum_{M\in\mathcal{M}}D_{k(x,M)}$.

  ``If'' direction is easy. Let $D(x,M)=D_{k(x,M)}$, then the first
  requirement is satisfied naturally. The second requirement is
  satisfied by the fact that $k(x,M)=k(x\oplus S,M)$ for any $x$, $M$
  and $S\in M$.

  Thus we now assume $H$ is $2$-decomposible, i.e, for every
  $x\in \set{0,1}^m$ and $M\in\mathcal{M}$, there exists $D(x,M)\ge 0$
  such that
  \begin{enumerate}
  \item $H(x)=\sum_{M\in\mathcal{M}}D(x,M)$, and
  \item $D(x,M)=D(x\oplus S,M)$ for every $S\in M$.
  \end{enumerate}
  We need to show that there exists $D_k\ge 0$ such that for every
  $x\in \set{0,1}^m$, $H(x)=\sum_{M\in\mathcal{M}}D_{k(x,M)}$.

  Let $\sigma\in S_m$ be a permutation on $[m]$. For every
  $x\in\set{0,1}^n$, we use $x_\sigma$ to denote
  $(x_{\sigma(1)},\dots,x_{\sigma(m)})$ and for every
  $M\in\mathcal{M}$, we use $M_{\sigma}$ to denote the partition on
  $[m]$ that
  $(x_i,x_j)\in M\iff (x_{\sigma(i)},x_{\sigma(j)})\in M_{\sigma}$. It
  is easy to see that for every $0\le k\le n$ and $\sigma\in S_m$,
  $(x,M)\in \Delta_k\iff (x_\sigma,M_\sigma)\in\Delta_k$.

  For every $k\ge 0$, we fix some $(x^{(k)},M^{(k)})\in \Delta_k$ and
  define
  $D_k= \frac{1}{m!}\sum_{\sigma\in S_m}
  D(x_\sigma^{(k)},M_\sigma^{(k)})$.
  An important fact is that the value of $D_k$ is an invariant for
  different choice of $(x^{(k)},M^{(k)})\in \Delta_k$. To see this,
  consider two pairs $(x,M),(x',M')\in \Delta_k$ where
  $x=(x_1,x_2,\dots,x_m)$ and $x'=(x_1',x_2',\dots,x_m')$, we aim to
  show that
  \begin{equation}
    \label{eq:sym}
    \sum_{\sigma\in S_m}D(x_\sigma,M_\sigma)=\sum_{\sigma\in S_m}D(x'_\sigma,M'_\sigma).
  \end{equation}

  We can assume without lost of generality that no pair
  $S=(x_i,x_j)\in M$ with $x_i=x_j=1$ and no pair
  $S'=(x_i',x_j')\in M'$ with $x_i'=x_j'=1$. This is because for every
  $S\in M$, the mapping
  $g((x_\sigma,M_\sigma))=((x\oplus S)_\sigma,M_\sigma)$ is a
  bijection between $\set{(x_\sigma,M_\sigma)\mid \sigma\in S_m}$ and
  $\set{((x\oplus S)_\sigma,M_\sigma)\mid \sigma\in S_m}$, and
  moreover $D(x_\sigma,M_\sigma)=D((x\oplus S)_\sigma,M_\sigma)$. Thus
  for every $S=(x_i,x_j)\in M$ with $x_i=x_j=1$, the identity
  \eqref{eq:sym} is equivalent if we replace $x$ by $x\oplus S$. The
  same argument holds for $x'$.

  Under this assumption, we have $\sum_{i=1}^nx_i=\sum_{i=1}^nx_i'$
  and both pairs belong to $\Delta_k$. This implies for some
  permutation $\pi\in S_m$, it holds that $(x_\pi,M_\pi)=(x',M')$ and
  justify \eqref{eq:sym}.

  It remains to verify that for every $x\in\set{0,1}^m$,
  $H(x)=\sum_{M\in\mathcal{M}}D_{k(x,M)}$. Since $H(\cdot)$ is
  symmetric, we have

  \begin{align*}
    H(x)
    &=\frac{1}{m!}\sum_{\sigma\in S_m}H(x_\sigma)
      =\frac{1}{m!}\sum_{\sigma\in S_m}\sum_{M\in\mathcal{M}}D(x_\sigma,M)
    =\frac{1}{m!}\sum_{M\in\mathcal{M}}\sum_{\sigma\in S_m}D(x_\sigma,M_\sigma)\\
    &=\frac{1}{m!}\sum_{k=0}^n\sum_{M\in\mathcal{M}:(x,M)\in \Delta_k}\sum_{\sigma\in S_m}D(x_\sigma,M_\sigma).
  \end{align*}
  It then follows from our discussion in the last paragraph that
  \begin{align*}
    H(x)=\sum_{k=0}^n\sum_{M\in\mathcal{M}:(x,M)\in \Delta_k}D_k
    =\sum_{M\in\mathcal{M}}D_{k(x,M)}.
  \end{align*}

  Therefore, the function $H$ is $2$-decomposible if and only if there
  exist $D_k\ge 0$ for every $k=0,1,\dots,n$ such that for every
  $x=(x_1,x_2,\dots,x_m)\in\set{0,1}^m$,
  \begin{align}
    \notag
    H(x)&=\sum_{M\in\mathcal{M}}D_{k(x,M)}=\sum_{k=0}^n\sum_{M\in\mathcal{M}:k(x,M)=k}D_k\\
        \label{eq:lineareq}
        &=\sum_{k=0}^n\abs{\set{M\in\mathcal{M}\mid
        k(x,M)=k}} D_k.
  \end{align}

  Since $H(\cdot)$ is a symmetric function, for every
  $x,x'\in\set{0,1}^m$ with same Hamming weight, identity
  \eqref{eq:lineareq} are the same. Moreover, the identity
  \eqref{eq:lineareq} for $x$ with Hamming weight $i$ is the same as
  the identity \eqref{eq:lineareq} for $x$ with Hamming weight $m-i$.
  For $i=|x|$, the identity \eqref{eq:lineareq} becomes
  \begin{equation*}
    h_i=\sum_{k=0}^n\abs{\set{M\in\mathcal{M}\mid
        k(x,M)=k}} D_k = \sum_{k=0}^n a_{ik} D_k,
  \end{equation*}
  where the second equality uses the (combinatorial) definition of
  $a_{ik}$.  Therefore, these $D_k\ge 0$ are the solution of the
  linear system $\mathbf{A}_m\mathbf{x}=\mathbf{h}$ defined in the
  statement of the lemma. This completes the proof for the case that
  $m$ is even.

  Then we consider the case that $m=2n+1$ is odd. Let $\mathcal{M}$
  denote the set of all partitions of $[m]$ into pairs and a
  singleton. The proof is similar to the case that $m$ is even, with
  some slight difference on verifying \eqref{eq:sym}, as we have to
  deal with the singleton in each $M\in\mathcal{M}$. We define an
  equivalent relation $\sim$ as that $(x,M)\sim(x',M')$ if
  $k(x,M)=k(x',M')$.  This definition is the same as the $m=2n$ case
  as the singleton plays no role.  For every $k=0,\dots,n$, we also
  define $\Delta_k=\set{(x,M)\mid k(x,M)=k}$ and claim the the
  function $H$ is $2$-decomposible if and only if for every
  $0\le k\le n$, there exists $D_k\ge 0$ such that for every
  $x\in\set{0,1}^m$, $H(x)=\sum_{M\in\mathcal{M}}D_{k(x,M)}$. The
  proof for the claim is almost identical as the even case. When
  verifying \eqref{eq:sym}, we can assume no pair $(x_i,x_j)\in M$
  with $x_i=x_j=1$ and that the singleton $(x_i)\in M$ satisfies
  $x_i=0$ (and the same assumption for $(x',M')$), then the remaining
  argument can go through.
%
\end{proof}

Our characterization of the windability of symmetric functions is
summarized by following theorem:

\begin{theorem}
  \label{thm:main}
  Given a symmetric function $F:\set{0,1}^d\to\mathbb{R}^+$, $F$ is
  windable if and only if for every pinning $G$ of $F$ with arity $m$,
  the function $H(x)=[h_0,h_1,\dots,h_m]\defeq G(x)G(\bar x)$
  satisfies the following condition: The linear equations
  $\mathbf{A}_m\mathbf{x}=\mathbf{h}$ has a nonnegative solution
  $\mathbf{x}\ge 0$, where
  $ \mathbf{h}=[h_0,h_1,\dots h_{\lfloor\frac{m}{2}\rfloor}]$.
\end{theorem}

We note that there exists an unique solution for
$\mathbf{A}_m\mathbf{x}=\mathbf{h}$ as $\mathbf{A}_m$ is a lower
triangular matrix. So we only need to check that this solution is
nonnegative or not.

\subsection{Properties of $\mathbf{A}_m$}
In this subsection, we obtain some properties of the matrix
$\mathbf{A}_m$ which are useful to verify that the linear equations
$\mathbf{A}_{m}\cdot \mathbf{x}= \mathbf{h}$ has a nonnegative
solution or not.

First of all, for all $i=0,1,\cdots, \lfloor\frac{m}{2}\rfloor$ we
have
$$\sum_{0\leq j\leq i}a_{ij}=(2\lfloor\frac{m-1}{2}\rfloor+1)!!=a_{00}.$$
This has a simple combinatorial explanation since the sum is the total
number of partitions of $m$ different objects into pairs and at most
one singleton. This implies the following lemma.
\begin{lemma}
  \label{lm:decom0}
  Let $m\geq 1$ and $c\geq0$,
  $\mathbf{A}_m \mathbf{x}=c\cdot\mathbf{1}$ has a nonnegative
  solution $\mathbf{x}=\frac{c}{a_{00}}\cdot \mathbf{1}$.
\end{lemma}

In the case that $m=2n$ is even, the matrix $\mathbf{A}_m$ has
non-zero entries $a_{ij}$ only if $i\equiv j\pmod 2$. Thus the
existence of nonnegative solution for the linear equations
$\mathbf{A}_m\mathbf{x}=\mathbf{h}$ is equivalent to the existence of
nonnegative solutions for the two linear equations
$\mathbf{A}_m\mathbf{x}=\mathbf{h}_0$ and
$\mathbf{A}_m\mathbf{x}=\mathbf{h}_1$, where $\mathbf{h}_0$
(resp. $\mathbf{h}_1$) is obtained from $\mathbf{h}$ by setting
$h_i=0$ for all odd (resp. even) $i$. This fact implies the following
corollary:

\begin{corollary}\label{cor:2decom-even}
  Let $H(x)=G(x)G(\bar x)$ be a symmetric function with arity
  $m=2n$. Define functions $H_0,H_1$ as $H_0= H \cdot \even$ and
  $H_1= H \cdot \odd$.
  %
  Then $H$
  is $2$-decomposible
  if and only if both $H_0$ and $H_1$ are $2$-decomposible.
\end{corollary}

Combined with Lemma~\ref{lm:decom0}, we directly have the following
lemma.

\begin{lemma}
  \label{lm:decom1}
  If $m=2n$ is even, $\mathbf{A}_m \mathbf{x}=\mathbf{h}$ has a
  nonnegative solution if $\mathbf{h}=\even$ or $\odd$.
\end{lemma}

The following lemma reveals an relation between $\mathbf{A}_{2n}$ and
$\mathbf{A}_{2n-1}$.
\begin{lemma}
  \label{lm:pdecom}
  Assume $n\geq 1$. Let
  $\mathbf{A}_{2n}=(a_{ij})\in \mathbb{R}^{(n+1)\times (n+1)}$, and
  $\mathbf{A}_{2n-1}=(a'_{ij})\in \mathbb{R}^{n\times n}$. If
  $0\leq i\leq n$ and $i\equiv j\pmod 2$, we have the following
  equality:
  \begin{equation}
    \label{eq:1}
    a_{ij}=a'_{i,j-1}+ a'_{ij} = a'_{i-1,j-1}+a'_{i-1,j}. \footnote{If $i=0$, the equality is $a_{00}=a'_{00}$. If $i=n$, the equality is $a_{nj}=a'_{n-1,j-1}+a'_{n-1,j}$.}
  \end{equation}
  Moreover, given two vectors
  $ \mathbf{h}\in \mathbb{R}^{(n+1)\times (n+1)}$ and
  $ \mathbf{h}'=\mathbb{R}^{n\times n}$, we have the following two
  properties:
  \begin{enumerate}
  \item If $ \mathbf{h}$ is odd (all even entries of $ \mathbf{h}$ are
    0), and $h'_{2i}=h'_{2i+1}=h_{2i+1}$ satisfies for
    $0\leq i\leq \lfloor n/2\rfloor$.\footnote{Since $h'_{n+1}$ does
      not exist, if $n$ is even and $i=\lfloor n/2\rfloor$, the
      condition is $h'_{n}=h_{n+1}$.} Then
    $\mathbf{A}_{2n-1}\cdot \mathbf{x'}= \mathbf{h}'$ has a
    nonnegative solution if and only if
    $\mathbf{A}_{2n}\cdot \mathbf{x}= \mathbf{h}$ has a nonnegative
    solution.
  \item If $ \mathbf{h}$ is even (all odd entries of $ \mathbf{h}$ are
    0), and $h'_{2i-1}=h'_{2i}=h_{2i}$ satisfies for
    $0\leq i\leq n/2$.\footnote{If $i=0$, the condition is
      $h'_0=h_0$.} Then
    $\mathbf{A}_{2n-1}\cdot \mathbf{x'}= \mathbf{h}'$ has a
    nonnegative solution if and only if
    $\mathbf{A}_{2n}\cdot \mathbf{x}= \mathbf{h}$ has a nonnegative
    solution.
  \end{enumerate}
\end{lemma}

\begin{proof}
  We first prove Equality~\ref{eq:1}. In fact, it is not hard to
  verify it by definition. Here we give a combinatorial explanation.
  Recall that $a_{ij}$ is the number of matchings in $\Delta_j$ when
  $\sum_{k\in [2n]} x_k=i$ ($0\leq i\leq n$). If $i\equiv j\pmod 2$
  and $i<n$, there must exist an entry of value 0. Assume that
  $x_{2n}=0$ without loss of generality. Then the matching among the
  remaining entries should be in either $\Delta_{j-1}$ or $\Delta_j$,
  and $\sum_{k\in [2n-1]}x_k=i$. Thus, we have
  $a_{ij}=a'_{i,j-1}+a'_{ij}$. Similarly, if $i\equiv j\pmod 2$ and
  $i>0$, there must exist an entry of value 1. We let $x_{2n}=1$
  without loss of generality. Then the matching among the remaining
  entries should be in either $\Delta_{j-1}$ or $\Delta_j$, and
  $\sum_{k\in [2n-1]}x_k=i-1$. In this case, we have that
  $a_{ij}=a'_{i-1,j-1}+a'_{i-1,j}$. Combine these two equalities, we
  prove Equality~\ref{eq:1}.

  If $ \mathbf{h}$ is odd, suppose $\mathbf{x}$ is the solution for
  the linear equations $\mathbf{A}_{2n}\cdot \mathbf{x}=
  \mathbf{h}$.
  Observe that $\mathbf{x}$ is also odd by the definition of
  $\mathbf{A}_{2n}$. Let $x'_{2i}=x'_{2i+1}=x_{2i+1}$ for
  $0\leq i\leq \lfloor n/2\rfloor$. We show that this $\mathbf{x'}$ is
  exactly the solution of
  $\mathbf{A}_{2n-1}\cdot \mathbf{x'}= \mathbf{h}'$. Then by the
  construction of $\mathbf{x'}$, we know that $\mathbf{x}$ is
  nonnegative if and only if $\mathbf{x'}$ is nonnegative, which
  completes the proof. Consider the $(2i)$th row
  ($0\leq i\leq \lfloor \frac{n-1}{2}\rfloor$) and $(2i+1)$th row of
  $\mathbf{A}_{2n-1}$ ($0\leq i\leq \lfloor \frac{n-2}{2}\rfloor$), we
  have the following equalities which shows that
  $\mathbf{A}_{2n-1}\cdot \mathbf{x'}= \mathbf{h}'$.
  \begin{align*}
    \sum_{0\leq j\leq 2i} a'_{2i,j}x'_j 
    &= \sum_{0\leq j\leq i} a'_{2i,2j}x'_{2j}+ a'_{2i,2j+1}x'_{2j+1} \\
    &=\sum_{0\leq j\leq i} a'_{2i,2j}x_{2j+1}+a'_{2i,2j+1}x_{2j+1} \\
    &= \sum_{0\leq j\leq i} (a'_{2i,2j}+a'_{2i,2j+1})x_{2j+1}\\
    &= \sum_{0\leq j\leq i} a_{2i+1,2j+1}x_{2j+1} = h_{2j+1} = h'_{2j},
  \end{align*}
  \begin{align*}
    \sum_{0\leq j\leq 2i+1} a'_{2i+1,j}x'_j
    &= \sum_{0\leq j\leq i} a'_{2i+1,2j}x'_{2j}+ a'_{2i+1,2j+1}x'_{2j+1} \\
    &=\sum_{0\leq j\leq i}  a'_{2i+1,2j}x_{2j+1}+a'_{2i+1,2j+1}x_{2j+1}\\
    &= \sum_{0\leq j\leq i} (a'_{2i+1,2j}+a'_{2i+1,2j+1})x_{2j+1}\\
    &= \sum_{0\leq j\leq i} a_{2i+1,2j+1}x_{2j+1} = h_{2j+1} = h'_{2j+1}.
  \end{align*}

  If $ \mathbf{h}$ is even, suppose $\mathbf{x}$ is the solution for
  the linear equations $\mathbf{A}_{2n}\cdot \mathbf{x}=
  \mathbf{h}$.
  Observe that $\mathbf{x}$ is also even. Let
  $x'_{2i-1}=x'_{2i}=x_{2i}$ for $0\leq i\leq \lfloor n/2\rfloor$. By
  the same argument as above, this $\mathbf{x'}$ is exactly the
  solution of $\mathbf{A}_{2n-1}\cdot \mathbf{x'}= \mathbf{h}'$. So we
  prove the whole lemma.
\end{proof}


\section{Counting $b$-Edge-Covers}
\label{sec:edgecover}
In this section, we obtain \textbf{FPRAS} for counting $b$-edge-cover
for $b\leq2$ as an application of our characterization. By
Theorem~\ref{thm:rapidmixing}, we need to prove that the function
$\geq b$ is windable for $b\leq2$, and bound the ratio of $Z_2/Z_0$.

\begin{lemma} 
  \label{lm:windp2}
  If $b\leq 2$, the weight functions $\geq b$ are windable.
\end{lemma}

\begin{lemma}
  \label{lm:ratio2}
  For any counting $b$-edge-cover instance, we have that
  $Z_2/Z_0\leq 4n^2$, where $n$ is the number of edges.
\end{lemma}

We first prove Lemma~\ref{lm:windp2}.  Consider the pinning function
$G$ of $\geq b$. Since $b\leq 2$, $G$ might be $\mathbf{1}_m$,
$(\geq 1)_m$ or $(\geq 2)_m$. Let
$H(x)=[h_0,h_1,\dots,h_m]\defeq G(x)G(\bar x)$, and let
$\mathbf{h}=[h_0,h_1,\dots h_{\lfloor\frac{m}{2}\rfloor}]$. By the
definition, we know that $\mathbf{h}$ can only be
$\mathbf{1}_{\lfloor\frac{m}{2}\rfloor}$,
$(\geq 1)_{\lfloor\frac{m}{2}\rfloor}$ or
$(\geq 2)_{\lfloor\frac{m}{2}\rfloor}$.  Then by
Theorem~\ref{thm:main}, we need to show that
$\mathbf{A}_m \mathbf{x}=\mathbf{h}$ always has a nonnegative
solution. Thus, we only need to prove the following lemma.

\begin{lemma}
  \label{lm:nnsol}
  If $m\geq 1$ and $b\leq 2$, $\mathbf{A}_m \mathbf{x}=(\geq b)$ has a
  nonnegative solution.
\end{lemma}

\begin{proof}
  If $b=0$, $\mathbf{h}=\mathbf{1}_n$ has been proved in
  Lemma~\ref{lm:decom0}. We assume $b=1,2$ in the following.  We
  consider two different cases: $m$ is even and $m$ is odd.

  The first case is that $m=2n$ is even.  If $b=1$,
  $\mathbf{h}=(\geq 1)_n$. By Corollary~\ref{cor:2decom-even}, we only
  need to prove that both $\mathbf{A}_m\mathbf{x}=\mathbf{h}_0$ and
  $\mathbf{A}_m\mathbf{x}=\mathbf{h}_1$ have nonnegative
  solutions. Observe that $\mathbf{h}_1=\odd_n$. By
  Lemma~\ref{lm:decom1}, we only need to consider
  $\mathbf{h}_0=(\geq 2)_n\cdot \even_n$. Let
  $x_{2j}=\tuple{1-(-1)^j\frac{(2j-1)!!}{\prod_{i=1}^{j}(2n-2i)}}\frac{1}{(2n-1)!!}$
  if $0 \leq j\leq \lfloor\frac{n}{2}\rfloor$ and $x_{j}=0$ if
  otherwise. Note that $x_0=1-\frac{(-1)!!}{1}=0$. If $j>0$, the
  numerator $(2j-1)!!$ is no larger than the denominator
  $\prod_{i=1}^{j}(2n-2i)$ because we have $2j-1\leq n-1\leq 2n-2$. So
  $x_{2j}\geq 0$ always holds. Thus, we prove that $\mathbf{x}$ is a
  nonnegative vector.

  The remaining task is to show $\mathbf{x}$ is the solution.  We note
  that $x_0=0$ and thus the first equation is satisfied. For $i$ is
  odd, it is easy to see that $\sum_{0\leq j\leq i}a_{ij} x_{j}=0$
  since we have $a_{ij}=0$ for even $j$ and $x_j=0$ for odd $j$. In
  the following, we only need to verify that
  $\sum_{0\leq j\leq i}a_{ij} x_{j}=1$ for even $i=2k\geq 2$. For
  these, we have
  \begin{align*}
    \sum_{0\leq j\leq k}a_{2k,2j}x_{2j}
    &=\sum_{0\leq j\leq k}a_{2k,2j} (\frac{1}{(2n-1)!!}+x_{2j}-\frac{1}{(2n-1)!!})\\
    &\overset{(\heartsuit)}{=} 1+\sum_{0\leq j\leq k}a_{2k,2j}(x_{2j}-\frac{1}{(2n-1)!!}) \\
    &= 1+\sum_{0\leq j\leq k} \binom{2k}{2j} \binom{2n-2k}{2j} (2j)!(2k-2j-1)!! 
    \cdot(2n-2k-2j-1)!!\tuple{x_{2j}-\frac{1}{(2n-1)!!}} \\
    &= 1-\sum_{0\leq j\leq k} (-1)^j \frac{(2k)!(2n-2k)!(n-j-1)!}{2(k-j)!(n-k-j)!j!(2n-1)!}\\
    &= 1-\frac{(2k)!(2n-2k)!}{2(2n-1)!}\sum_{0\leq j\leq k} (-1)^j \frac{(n-j-1)!}{(k-j)!(n-k-j)!j!} \\
    &\overset{(\diamondsuit)}{=} 1-\frac{(2k)!(2n-2k)!}{2(2n-1)!}\sum_{0\leq j\leq k} \frac{(-1)^j {k \choose j}{n-j \choose k}}{n-j} \\
    &= 1-0 = 1,
  \end{align*}
  where $(\heartsuit)$ is because the sum of entries in each row of
  $\mathbf{A}_m$ is $(2n-1)!!$, which equals to the total number of
  partitions. The equality $(\diamondsuit)$ uses the fact
  $\sum_{0\leq j\leq k} \frac{(-1)^j {k \choose j}{n-j \choose
      k}}{n-j}=0$, which is by the following technical Lemma.

  \begin{lemma}
    \label{lm:com}
    $\sum_{j=0}^m \frac{(-1)^j {m \choose j}{n-j \choose m}}{n-j}=0$.
  \end{lemma}

\begin{proof}
  Consider
  $f(x)=\sum_{j=0}^m \frac{(-1)^j {m \choose j}{n-j \choose m}x^j
  }{n-j}$. It is not hard to see that
$$f(x)=\frac{{n\choose m} {}_2F_1(-m,m-n;1-n;x)}{n},$$
where
${}_2F_1(a,b;c;z)=\sum_{i\geq 0}\frac{(a)_i (b)_i}{(c)_i}\cdot
\frac{z^i}{i!}$. Here $(a)_i=\prod_{j=0}^{i-1} (a+j)$.

By Equality 15.3.3 in~\cite{abramowitz1965handbook},
${}_2F_1(-m,m-n;1-n;x)=(1-x)\cdot {}_2F_1(1+m-n,1-m;1-n;x)$. Let
$x=1$, we prove the lemma.
\end{proof}

If $b=2$, $\mathbf{h}=(\geq 2)_n$. We still consider the linear
equations $\mathbf{A}_m\mathbf{x}=\mathbf{h}_0$ and
$\mathbf{A}_m\mathbf{x}=\mathbf{h}_1$. Note that
$\mathbf{h}_0=(\geq 2)_n\cdot \even_n$ which has been proved in the
last case. So we focus on $\mathbf{h}_1$ which equals to
$(\geq 3)_n\cdot \odd_n$. Let
$x_{2j+1}=(1-(-1)^j\frac{(2j+1)!!}{\prod_{i=2}^{j+1}(2n-2i)})\frac{1}{(2n-1)!!}$
($0\leq j\leq \frac{n-1}{2}$). Otherwise let $x_{j}=0$. Then we show
the correctness.

If $j=0$, note that $x_1=0$. If $j=1$, it is not hard to see that
$x_3>0$. If $j>1$, we have $n\geq 5$. Observe that the numerator
$(2j+1)!!$ is no larger than the denominator
$\prod_{i=2}^{j+1}(2n-2i)$ since we have $2j+1\leq 2n-4$. So
$x_{2j+1}\geq 0$ always holds. Thus, we prove that $\mathbf{x}$ is a
non-negative vector.

The remaining task is to show that $\mathbf{x}$ is exactly the
solution. We note that $x_1=0$ and thus the second equation is
satisfied. For $i$ is even, it is easy to see that
$\sum_{0\leq j\leq i}a_{ij} x_{j}=0$ since we have $a_{ij}=0$ for odd
$j$ and $x_j=0$ for even $j$. In the following, we only need to
consider the $(2k+1)$th rows
($0\leq k\leq \lfloor\frac{n-1}{2}\rfloor$). In fact, we have the
following equalities.
\begin{align*}
  &\sum_{0\leq j\leq k}a_{2k+1,2j+1} x_{2j+1}\\
  =&\;\sum_{0\leq j\leq k} a_{2k+1,2j+1}\tuple{\frac{1}{(2n-1)!!}+x_{2j+1}-\frac{1}{(2n-1)!!}} 
  \overset{(\heartsuit)}{=} 1+ \sum_{0\leq j\leq k} a_{2k+1,2j+1}\tuple{x_{2j+1}-\frac{1}{(2n-1)!!}}\\
  =&\; 1+\sum_{0\leq j\leq k} \binom{2k+1}{2j+1} \binom{2n-2k-1}{2j+1} 
  \cdot(2j+1)!(2k-2j-1)!! (2n-2k-2j-3)!!
  \cdot\tuple{x_{2j+1}-\frac{1}{(2n-1)!!}} \\
  =&\; 1-\frac{(2k+1)!(2n-2k-1)!(n-1)}{(2n-1)!}
  \cdot\sum_{0\leq j\leq k} (-1)^j \frac{(n-j-2)!}{(k-j)!(n-k-j-1)!j!} \\
  \overset{(\diamondsuit)}{=}&\; 1-\frac{(2k+1)!(2n-2k-1)!(n-1)}{(2n-1)!}
  \cdot\sum_{0\leq j\leq k} \frac{(-1)^j {k \choose j}{n-1-j \choose k}}{n-1-j}\\
  =&\; 1-0 = 1.
\end{align*}
where $(\heartsuit)$ is because the sum of entries in each row of
$\mathbf{A}_m$ is $(2n-1)!!$ which equals to the total number of
partitions, and $(\diamondsuit)$ uses Lemma~\ref{lm:com}.

If $m=2n-1$ is odd. We want to show that
$\mathbf{A}_{2n-1}\cdot \mathbf{x}=(\geq b)_{n-1}$ has a nonnegative
solution for $b\leq 2$.  If $b=1$, by Lemma~\ref{lm:pdecom}, we only
need to show that
$\mathbf{A}_{2n}\cdot \mathbf{x}=(\geq 1)_n\cdot \even_n=(\geq
2)_n\cdot \even_n$
has a non-negative solution, which has been proved in the first
case. Finally, if $b=2$, by Lemma~\ref{lm:pdecom}, we only need to
prove that $\mathbf{A}_{2n}\cdot \mathbf{x}=(\geq 2)_n \cdot \odd $
has a nonnegative solution. Note that
$(\geq 2)_n \cdot \odd = (\geq 3)_n \cdot \odd_n$. By the first case,
we finish the proof.

Thus, we prove that $\mathbf{A}_m \mathbf{x}=(\geq b)$ always has a
nonnegative solution if $b\leq 2$.
\end{proof}

The second part is to prove Lemma~\ref{lm:ratio2}.

\begin{proof}
  We construct a mapping from $\Omega_2$ to $\Omega_0$ to bound
  $Z_2/Z_0$.  For any satisfying assignment $x\in \{0,1\}^{2n}$ in
  $\Omega_2$, assume that $i,j$ are the two half edges which violates
  the equality constraint on edges, and $x_i = x_j =0$ (the
  corresponding other two half edges are assigned $1$). Let $y$ be the
  assignment obtained by $x$ flipping on $i$th and $j$th entries. Note
  that $y\in \Omega_0$ is also a satisfying assignment by the
  definition of $b$-edge-cover.

  On the other hand, from a satisfying assignment $y\in \Omega_0$, we
  can construct at most $4n^2$ satisfying assignments $x\in \Omega_2$
  by flipping on two half edges. So we map at most $4n^2$ satisfying
  assignments $x\in \Omega_2$ to $y$. Thus, we have $Z_2/Z_0\leq 4n^2$
  by this mapping.
\end{proof}

Combining Lemma~\ref{lm:windp2} and~\ref{lm:ratio2}, we have the
following theorem.

\begin{theorem}
  \label{thm:unbedgecover}
  There is an \textbf{FPRAS} for counting $b$-edge-cover problems if
  $b\leq 2$.
\end{theorem}

\section{Counting $b$-Matchings}
In this section, we provide another application for counting $b$-matchings.
\begin{theorem}
\label{thm:unbmatching}
There is an \textbf{FPRAS} for counting $b$-matching problems if $b\leq 7$.
\end{theorem}
Similarly, by Theorem~\ref{thm:rapidmixing}, we only need to prove the following two lemmas.

\begin{lemma}
\label{lm:windp}
If $b\leq 7$, the weight functions $\leq b$ are windable.
\end{lemma}

\begin{lemma}
\label{lm:ratio}
For any counting $b$-matching instance, we have that $Z_2/Z_0\leq 4n^2$, where $n$ is the number of edges.
\end{lemma}

For preparation, we show the following lemma first.

\begin{lemma}
\label{lm:decom2}
Let $n=\lfloor\frac{m}{2}\rfloor$. Then $\mathbf{A}_m\mathbf{x}=(=n)_n$ has a nonnegative solution.
\end{lemma}

\begin{proof}
Since the RHS only has one non-zero entry at the last row, it is easy to see that $x_n=\frac{1}{a_{nn}}$ and  $x_i=0$ for $i=0,1,\cdots, n-1$ is a non-negative solution.
%
%
\end{proof}

Now we are ready to prove Lemma~\ref{lm:windp}.

\begin{proof} (Lemma~\ref{lm:windp})
Consider the pinning function $G$ of $\leq b$. We have that $G=(\leq k)_m$, where $k\leq 7$. Recall that we define $H(x)=[h_0,h_1,\dots,h_m]\defeq G(x)G(\bar x)$. Then we have $H=[m-k,k]_m$. To make $H$ non-trivial, we need $k \leq m \leq 2 k$.
Let $\mathbf{h}=[h_0,h_1,\dots h_{\lfloor\frac{m}{2}\rfloor}]$, then $h=(\geq m-k)_{\lfloor\frac{m}{2}\rfloor}$.
%
%
If $m \leq k+2$, then $\mathbf{h}=(\geq l)$ with $l\leq 2$ which has been proved by Lemma~\ref{lm:nnsol}.
By Lemma~\ref{lm:decom2}, the cases that $m=2k$ and $m=2k-1$ are also correct. So we only need to consider the cases that $k+3\leq n\leq 2k-2$ and $k\leq 7$. We enumerate all of them in the following\\
\textbf{Case $k=5,m=8$.}  $\mathbf{x}=(0,0,0,\frac{1}{60},\frac{1}{24})$ is the non-negative solution.\\
\textbf{Case $k=6,m=9$.} $\mathbf{x}=(0,0,0,\frac{1}{360},\frac{1}{360})$ is the non-negative solution.\\
\textbf{Case $k=6,m=10$.}  $\mathbf{x}=(0,0,0,0,\frac{1}{360},\frac{1}{120})$ is the non-negative solution.\\
\textbf{Case $k=7,m=10$.}  $\mathbf{x}=(0,0,0,\frac{1}{630},\frac{1}{360},\frac{1}{2520})$ is the non-negative solution.\\
\textbf{Case $k=7,m=11$.} $\mathbf{x}=(0,0,0,0,\frac{1}{2520},\frac{1}{2520})$ is the non-negative solution. \\
\textbf{Case $k=7,m=12$.}   $\mathbf{x}=(0,0,0,0,0,\frac{1}{2520},\frac{1}{720})$ is the non-negative solution.\\
\end{proof}

The remaining task is to prove Lemma~\ref{lm:ratio}.

\begin{proof} (Lemma~\ref{lm:ratio})
The argument is almost the same as Lemma~\ref{lm:ratio2} except that from a satisfying assignment $x\in \Omega_2$, we map it to a satisfying assignment $y\in \Omega_0$ by deleting two half edges, instead of adding two half edges. Again, we construct a mapping from $\Omega_2$ to $\Omega_0$, and show that $Z_2/Z_0\leq 4n^2$.
\end{proof}


{\noindent \bf Remark:} Our FPRAS for both $b$-matchings and $b$-edge-covers can be extended to instances with edge weights.  On the other hand, the results cannot be extended to counting $8$-matchings or $3$-edge-covers since these constraint functions are not windable. These facts are also showed  by our characterization theorem and we present them in the following two sections.

\section{Edge Weighted $b$-Edge-Covers and $b$-Matchings }

In this section, we consider the version that each edge $e\in E$ has a nonnegative weight $w_e$. We want to show that both counting weighted $b$-edge-cover and $b$-matching problems have an \textbf{FPRAS}. 

Given a graph $G=(V,E)$. The trick is to add a constraint on each edge. For each edge $e$, we separate it into two edges $e^0$ and $e^1$. Between $e^0$ and $e^1$, we add a new constraint $(1,0,w_e)$. Now we construct a new graph $G'=(V\cup E, E^0\cup E^1)$. 
It is easy to see that the partition function for this new Holant instance is exactly the partition function for the edge weighted counting problem.

We first prove the constraint for each edge is windable.

\begin{lemma}
\label{lm:edgew}
If $a\geq 0$, the function $(1,0,a)$ is windable.
\end{lemma}

\begin{proof}
For all pinnings $G$ of this function, we can observe that $G\overline{G}$ is either $\mathbf{0}$ or $c\cdot \mathbf{1}$, where $c$ is some nonnegative constant. By Lemma~\ref{lm:windto2} and~\ref{lm:decom0}, we prove the lemma .
\end{proof}

Compared to the unweighted version, we have $\abs{E}$ more constraints on edges. Note that the half edges are between vertex constraints and edge constraints. In other words, each edge $e\in E$ is partitioned into four half edges. It only needs to show that $Z_2/Z_0$ is still bounded. We first consider the weighted $b$-edge-cover problems.

\begin{lemma}
\label{lm:wratio2}
For any counting $b$-edge-cover instance where $b\leq 2$, we have that $Z_2/Z_0\leq \frac{16n^2}{\min w_e^2}$.\footnote{We assume $\min_e w_e$ is a constant. This assumption is reasonable. Since if $\min_e w_e$ is exponentially small, counting weighted $b$-edge-cover problem can be as hard as minimal edge-cover problem.} Here, $n$ is the number of edges.
\end{lemma}

\begin{proof}
Similar to Lemma~\ref{lm:ratio2}, we construct a mapping from $\Omega_2$ to $\Omega_0$. Since the half edges are different, the rules for the mapping are also different.

Consider a satisfying assignment $x\in \{0,1\}^{2n}$ in $\Omega_2$, exactly two pair of half edges disagree with each other. We call them 'bad' pairs. For an edge $e$, we partition it into four different half edges. If there exists a 'bad' pair of half edges on $e$, there might be exactly one, two or three half edges of value 1. We call this edge a 'bad' edge. Note that there are at most two such 'bad' edges. Assume they are $e_1$ and $e_2$.  Let $y$ be the assignment obtained by $x$ fixing all half edges to be 1 on $e_1$ and $e_2$. Note that $y\in \Omega_0$ is also a satisfying assignment by the definition of $b$-edge-cover. Moreover, $F(x)/F(y)\leq \frac{1}{\min_e w_e^2}$.

On the other hand, from a satisfying assignment $y\in \Omega_0$, we can construct at most $16n^2$ satisfying configurations $x\in \Omega_2$ by flipping on two random half edges. Note that for each such $x$, we also have $F(x)/F(y)\leq \frac{1}{\min_e w_e^2}$. Moreover, we map at most $16n^2$ satisfying configurations $x\in \Omega_2$ to $y$. Thus, by this mapping, we have that $Z_2/Z_0\leq \frac{16n^2}{\min_e w_e^2}$.
\end{proof}

\begin{theorem}
\label{thm:bedgecover}
There is an \textbf{FPRAS} for counting weighted $b$-edge-cover problems if $b\leq 2$.
\end{theorem}

For counting $b$-matching problems, we have similar results.

\begin{lemma}
\label{lm:wratio}
For any counting weighted $b$-matching instance where $b\leq 7$, we have that $Z_2/Z_0\leq 16n^2\max_e w_e^2$. \footnote{In this paper, we assume $\max_e w_e$ is a constant. This assumption is reasonable. Since if $\max_e w_e$ is exponentially large, counting weighted $b$-matching problem can be as hard as counting perfect matching.} Here, $n$ is the number of edges.
\end{lemma}

\begin{proof}
The proof is very similar to Lemma~\ref{lm:ratio2}, except that from a satisfying assignment $x\in \Omega_2$, we map $x$ to an assignment $y\in \Omega_0$ by fixing all half edges to be 1 instead of 0 on ``bad'' edges. Another difference is that we have $F(x)/F(y)\leq \max_e w_e^2$.
\end{proof}

Combined with Theorem~\ref{thm:rapidmixing}, we have the following theorem.

\begin{theorem}
\label{thm:bmatching}
There is an \textbf{FPRAS} for counting weighted $b$-matching problems if $b\leq 7$.
\end{theorem}

\noindent
\textbf{Remark:} Observe the weight function $H=(1,0,w_e)$. Note that the even entries of $H$ is a geometric sequence. In general, we have the following lemma.

\begin{lemma}
\label{lm:geo}
For a symmetric function $H:\{0,1\}^J\rightarrow \mathbb{R}^+$, if both the even and the odd subsequences are geometric sequences, then $H$ is a windable function.
\end{lemma}

\begin{proof}
We still focus on showing that for each pinning $G: \{0,1\}^m\rightarrow \mathbb{R}^+$ of $H$, $\mathbf{A}_m\mathbf{x}=\mathbf{h}$ has a nonnegative solution by Theorem~\ref{thm:main}, where $\mathbf{h}$ is the prefix of $G\cdot \overline{G}$.

If $m$ is odd, by the property of geometric sequences, we observe that $h=c\cdot \mathbf{1}$ ($c>0$). If $m=2n$ is even, by Corollary~\ref{cor:2decom-even}, we only need to show that both $\mathbf{A}_m\mathbf{x}=\mathbf{h}_0$ and $\mathbf{A}_m\mathbf{x}=\mathbf{h}_1$ have a nonnegative solution. By the property of geometric sequences, it is not hard to see that $h_0=c_1\cdot \even_n$ and $h_1=c_2\cdot \odd_n$ ($c_1,c_2>0$). By Lemma~\ref{lm:decom0} and~\ref{lm:decom1}, we prove that $H$ is windable.
\end{proof}

By Lemma~\ref{lm:geo}, we can show that FPRAS exists for this class of symmetric functions similar to $B$-matching. Note that $[1,\mu,1,\mu,\ldots]$ is a special case, which has a well-known FPRAS in~\cite{jerrum1993polynomial}. So we give an FPRAS for a more general class of counting problems.

\section{Unwindable Functions}
\label{app:unwind}

In this section, we give some examples of unwindable functions, which shows that our approach cannot be directly extended to  $3$-edge-cover and $8$-matching problems.
\begin{lemma}
\label{lm:windno}
If $b\geq 3$ and $|J|\geq b+8$, the weight functions $(\geq b)_J$ are not windable.
\end{lemma}

\begin{proof}
If $b\geq 3$ and $|J|\geq b+8$, there must be a pinning $G$ by $\p$, where $G=(\geq 3)_{11}$. By Theorem~\ref{thm:main}, we only need to show that $\mathbf{A}_{11}\cdot \mathbf{x}=(\geq 3)_6$ has nonpositive solution. In fact, we know that $\mathbf{x}=(0,0,0,\frac{1}{5040},\frac{1}{5040},-\frac{1}{10080})$ by calculation.
\end{proof}

Lemma~\ref{lm:windno} shows that why our technique can not work for arbitrary $b$-edge-covers. By this lemma, we can conclude the following corollary which shows that why winding technique does not work for arbitrary $b$-matchings.

\begin{corollary}
If $b\geq 8$ and $|J|\geq b+3$, the weight functions $(\leq b)_J$ are not windable.
\end{corollary}

\begin{proof}
For a weight function $F=(\leq b)_J$, let $F'=(\geq |J|-b)_J$. Consider a pinning $G$ of $F$ by $\p$. We construct another pinning $G'$ of $F'$ by $\overline{\p}$. Note that for any $x$, we have that $G(x)=G'(\overline{x})=\overline{G'}(x)$. Then $G\cdot \overline{G}$ is exactly the same as $G'\overline{G'}$. So $F$ is windable if and only if $F'$ is windable. \\
Note that $|J|-b\geq 3$ and $|J|\geq |J|-b+8$. By Lemma~\ref{lm:windno}, we prove the corollary.
\end{proof}

\bibliographystyle{plain}
\bibliography{refs}

\appendix










\section*{Appendix}

To be self-contained and for the convenience of readers, we include a formal proof for Theorem \ref{thm:rapidmixing} in this appendix.
These proofs are essentially adapted from~\cite{mcquillan2013approximating}. 


We first construct a Markov chain to sample from $\Omega_0\cup\Omega_2$.

Let $\Lambda=(G(V,E),(f_v)_{v\in V})$ be an instance with $\abs{V}=n$
and every $f_v$ is windable. Let $\mathcal{E}$ be the set of half
edges in $G$. The state space of the chain is
$\Omega=\Omega_0\cup \Omega_2$. For every two configuration
$\sigma,\pi\in\Omega$, the transition probability $P'(\sigma,\pi)$ is
defined as
\[
  P'(\sigma,\pi)=
  \begin{cases}
    \frac{2}{n^2}\min\tuple{1,\frac{w_\Lambda(\pi)}{w_\Lambda(\sigma)}}, &
     \mbox{if }d(\sigma,\pi)=2;\\
    1-\frac{2}{n^2}\sum_{\rho:d(\sigma,\rho)=2}\min\tuple{1,\frac{w_\Lambda(\rho)}{w_\Lambda(\sigma)}}, &
     \mbox{if }\sigma=\pi;\\
    0, & \mbox{otherwise,}
  \end{cases}
\]
where $d(\sigma,\pi)$ denote the Hamming distance between $\sigma$ and
$\pi$.

Our Markov chain is the lazy version of above, i.e., for every two
configurations $\sigma,\pi\in\Omega$, define
$P(\sigma,\pi)=\frac{1+P'(\sigma,\pi)}{2}$ if $\sigma=\pi$ and $P(\sigma,\pi)=\frac{P'(\sigma,\pi)}{2}$ if $\sigma\ne\pi$\footnote{Note that the chain defined here is slightly different with the one used in \cite{mcquillan2013approximating}}.

For every $\sigma\in\Omega$, we denote
$\mu_\Lambda(\sigma)\defeq\frac{w_\Lambda(\sigma)}{Z_0+Z_2}$ and for
every set $S\subseteq \Omega$, we denote
$\mu_\Lambda(S)\defeq\sum_{\sigma\in S}\mu_{\Lambda}(\sigma)$.

The following rapid mixing result for above chain was established in
\cite{mcquillan2013approximating}. For self-reducible instances, it is standard
to obtain \textbf{FPRAS} from this rapidly mixing Markov chain \cite{samp_JVV86}.

\begin{lemma}\label{lem:mixing}
  For all $\sigma\in\Omega$ and all non-negative integers $t$, we have
  \[
    \left\|P^t(\sigma,\cdot)-\mu_\Lambda\right\|_{TV} \le
    \frac{1}{2}\tuple{\mu_\Lambda(\sigma)}^{-\frac{1}{2}}\exp{-t\cdot\mu_\Lambda(\Omega_0)^2/n^4}.
  \]
\end{lemma}

The remaining part of this section is devoted to prove Lemma
\ref{lem:mixing}.

\section{Congestion and Canonical Paths}

Let $\mathcal{G}(\Omega,\mathcal{E})$ be the transition graph of our
Markov chain where for every pair of configurations
$\sigma,\pi\in \Omega$, $(\sigma,\pi)\in \mathcal{E}$ if and only if
$P(\sigma,\pi)>0$.

A \emph{flow-path} $\gamma$ is a directed path in $\mathcal{G}$
equipped with a weight $\wt{\gamma}$. Canonical paths $\Gamma$ from
$X\subseteq \Omega$ to $Y\subseteq\Omega$ is a set of flow-paths
satisfying
\[
  \sum_{\substack{\scriptsize \mbox{paths }\gamma\in\Gamma\\\mbox{\scriptsize~from $x$ to
      $y$}}}\wt{\gamma}=\pi(x)\pi(y)\quad\mbox{for all $x\in X$ and
    $y\in Y$}.
\]
The \emph{congestion} of $\Gamma$ is defined as
\[
  \rho(\Gamma)\defeq\max_{(\sigma,\pi)\in\mathcal{E}}\frac{1}{\pi(\sigma)P(\sigma,\pi)}\sum_{\gamma\in\Gamma\mbox{\scriptsize~s.t. }(\sigma,\pi)\in\gamma}\wt{\gamma}.
\]

The following lemma was established in \cite{diaconis1991geometric}
and \cite{sinclair1992improved}:

\begin{lemma}
  For every canonical paths $\Gamma$ from $\Omega$ to $\Omega$, every
  $\sigma\in\Omega$ and every nonegative $t$, it holds that
  \[
    \left\|P^t(\sigma,\cdot)-\mu_\Lambda(\cdot)\right|_{TV}\le
    \frac{1}{2}\tuple{\mu_\Lambda(\sigma)}^{-\frac{1}{2}}\exp{-\frac{t}{n\rho(\Gamma)}}.
  \]
\end{lemma}
Thus it remains to construct a flow-path $\Gamma$ such that
$\rho(\Gamma)\le \frac{n^3}{\mu_\Lambda(\Omega_0)^2}$.

\section{The Construction of Canonical Paths}

In this section, we describe the construction of canonical paths.

\bigskip
\paragraph{Flow from $\Omega_0$ to $\Omega$.}

Let $\sigma\in\Omega_0$ and $\pi\in\Omega_2$ be two configurations and
$z=\sigma\oplus\pi$. Consider a tuple
$\tuple{M_v\in \mathcal{M}_{z|_{\mathcal{E}(v)}}}_{v\in V}$, define
$T$ as the set of singletons in $\bigcup_{v\in V}M_v$, i.e.,
$T\defeq\set{S\in M_v\mid v\in V\mbox{ and }S\mbox{ is a singleton}}$.
We fix a partition of $T$ into pairs (note that $|T|$ is even by the
definition of $\Omega_0$ and $\Omega_2$) and denote the partition as
$M'$.  Define $M\defeq\bigcup_{v\in V}M_v\cup M'\in\mathcal{M}_z$, we
call $M$ the partition induced by
$\tuple{M_v\in \mathcal{M}_{z|_{\mathcal{E}(v)}}}_{v\in V}$.

Then for every tuple
$\tuple{M_v\in \mathcal{M}_{z|_{\mathcal{E}(v)}}}_{v\in V}$, we define
a canonical path $\gamma_{\sigma,\pi,M}$ as follows, where
$M\in\mathcal{M}_z$ is the partition induced by the tuple: We first
construct a graph $G_{M,z}=(V_z,E_M)$ where

\begin{itemize}
\item $V_z=\set{e_v\in \mathcal{E}\mid z(e_v)=1}$;
\item $E_M=M\cup \set{\set{e_u,e_v}\in V_z^2\mid \set{u,v}\in E}$.
\end{itemize}
Since both $\sigma,\pi\in\Omega$, which implies $G_{M,z}$ is a graph
consisting of disjoint cycles and a path.  We recursively choose an
order of edges $\set{e_1,e_2,\dots,e_m}$ in $E_M$ as follows:
\begin{itemize}
\item If there is a unique path $P=(e_1,e_2,\dots,e_k)$, then start
  from $e_1$ and choose edges along the path in the same order. After
  this is done, remove $P$.
\item If there is no path, choose a cycle $C=(e_1,e_2,\dots,e_k,e_1)$
  such that $\set{e_1,e_2}\in M$. Then start from $e_1$ and choose
  edges along the cycle. After this is done, remove $C$.
\end{itemize}
This order induces an order of pairs in $M$. We denote it by
$\set{S_1,S_2,\dots, S_t}$ where each $S_k\in M$ is a pair of half
edges.

For every $k=0,1,2,\dots,t$, let $E_k\defeq \bigcup_{i=1}^kS_k$.  We
then construct a flow-path $\gamma_{\sigma,\pi,M}$ in $\Omega$ as
\[
  \sigma=\sigma\oplus E_0\to \sigma\oplus E_1\to\dots\to\sigma\oplus
  E_t=\pi,
\]
and equip the path with weight
\[
  \wt{\gamma_{\sigma,\pi,M}}=\prod_{v\in
    V}B_v(\sigma|_{\mathcal{E}(v)},\pi|_{\mathcal{E}(v)},M_v)/(Z_0+Z_2)^2,
\]
where for every $v\in V$, $B_v(\cdot,\cdot,\cdot)$ is the set of
values witnessing $f_v$ is windable.

Then for every $\sigma\in \Omega_0$ and $\pi\in\Omega$, it holds that
\begin{align*}
  \sum_{M\in\mathcal{M}_z}\wt{\gamma_{\sigma,\pi,M}}
  &=\frac{1}{(Z_0+Z_2)^2}\sum_{\set{M_v\in \mathcal{M}_{z\cap \mathcal{E}(v)}}_{v\in V}}\prod_{v\in V}B_v(\sigma|_{\mathcal{E}(v)},\pi|_{\mathcal{E}_v},M_v)\\
  &=\frac{1}{(Z_0+Z_2)^2}\cdot\prod_{v\in V}\sum_{M_v\in\mathcal{M}_{z\cap\mathcal{E}(v)}}B_v(\sigma|_{\mathcal{E}(v)},\pi|_{\mathcal{E}(v)},M_v)\\
  &\overset{(\heartsuit)}{=}\frac{1}{(Z_0+Z_2)^2}\cdot\prod_{v\in V}f_v(\sigma|_{\mathcal{E}(v)})f_v(\pi|_{\mathcal{E}(v)})\\
  &=\mu_\Lambda(\sigma)\mu_{\Lambda}(\pi),
\end{align*}
where $(\heartsuit)$ is due to the definition of windability. We
denote $\Gamma_0$ the canonical paths constructed above.

\bigskip
\paragraph{Flow from $\Omega$ to $\Omega$.}

For every $\sigma,\pi\in\Omega$, for every $\rho\in\Omega_0$, every
$M_1\in\mathcal{M}_{\sigma\oplus\rho}$, every
$M_2\in\mathcal{M}_{\rho\oplus\pi}$, we construct a path
$\gamma_{\sigma,\pi,\rho,M_1,M_2}$ which is the concatenation of
$\gamma_{\sigma,\rho,M_1}$ and $\gamma_{\rho,\pi,M_2}$ (since the
transition graph of our Markov chain is undirected, we can safely
reverse paths in $\Gamma_0$). The weight of
$\gamma_{\sigma,\pi,\rho,M_1,M_2}$ is
$\frac{\wt{\gamma_{\sigma,\rho,M_1}}\wt{\gamma_{\rho,\pi,M_2}}}{\mu_\Lambda(\rho)\mu_\Lambda(\Omega_0)}$. The
flow is legal since
\begin{align*}
  \sum_{\rho\in\Omega_0}\sum_{M_1\in\mathcal{M}_{\sigma\oplus\rho}}\sum_{M_2\in\mathcal{M}_{\rho\oplus\pi}}\wt{\gamma_{\sigma,\pi,\rho,M_1,M_2}}
  &=\sum_{\rho\in\Omega_0}\sum_{M_1\in\mathcal{M}_{\sigma\oplus\rho}}\sum_{M_2\in\mathcal{M}_{\rho\oplus\pi}}\frac{\wt{\gamma_{\sigma,\rho,M_1}}\wt{\gamma_{\rho,\pi,M_2}}}{\mu_\Lambda(\rho)\mu_\Lambda(\Omega_0)}\\
  &=\sum_{\rho\in\Omega_0}\frac{\mu_\Lambda(\sigma)\mu_\Lambda(\rho)\mu_\Lambda(\pi)}{\mu_\Lambda(\Omega_0)}\\
  &=\mu_\Lambda(\sigma)\mu_\Lambda(\pi).
\end{align*}

\section{Analysis}

In this section, we bound the congestion of the canonical paths
constructed in the previous section.

\begin{lemma}\label{lem:Z2bound}
  Let $\Lambda=(G(V,E),(f_v)_{v\in V})$ be an instance with
  $\abs{V}=n$ and every $f_v$ is windable, then $Z_0Z_4\le Z_2Z_2$.
\end{lemma}
\begin{proof}
  Note that
  \begin{align*}
    Z_0Z_4
    &=\sum_{\substack{\sigma\in\Omega_0\\\pi\in\Omega_4}}w_\Lambda(\sigma)w_\Lambda(\pi)\\
    &=\sum_{\substack{\sigma\in\Omega_0\\\pi\in\Omega_4}}\prod_{v\in
    V}f_v(\sigma|_{\mathcal{E}(v)})f_v(\pi|_{\mathcal{E}(v)})\\
    &=\sum_{\substack{\sigma\in\Omega_0\\\pi\in\Omega_4}}\prod_{v\in
    V}\sum_{M_v\in\mathcal{M}_{z|_{\mathcal{E}(v)}}}B_v(\sigma|_{\mathcal{E}(v)},\pi|_{\mathcal{E}(v)},M_v)\\
    &=\sum_{\substack{\sigma\in\Omega_0\\\pi\in\Omega_4}}\sum_{\set{M_v\in\mathcal{M}_{z|_{\mathcal{E}(v)}}}_{v\in V}}\prod_{v\in
    V}B_v(\sigma|_{\mathcal{E}(v)},\pi|_{\mathcal{E}(v)},M_v),
  \end{align*}
  where in the last two lines $z=\sigma\oplus \pi$ and
  $B_v(\cdot,\cdot,\cdot)$ is the family of values witnessing the
  windability of $f_v$.

  Fix $(\sigma,\pi)\in\Omega_0\times\Omega_4$ and
  $\set{M_v\in\mathcal{M}_{z|_{\mathcal{E}(v)}}}_{v\in V}$ where
  $z=\sigma\oplus\pi$. Let $M$ be the set of pairs in
  $\bigcup_{v\in V}M_v$. Define a graph $G_{M,z}=(V_z,E_M)$ where
  \begin{itemize}
  \item $V_z=\set{e_v\in\mathcal{E}\mid z(e_v)=1}$;
  \item $E_M=M\cup\set{\set{e_u,e_v}\in V_z^2\mid \set{u,v}\in E}$.
  \end{itemize}
  Since $(\sigma,\pi)\in\Omega_0\times\Omega_4$, $G_{M,z}$ consists of
  two disjoint paths and many disjoint cycles. Let $P$ be one of the
  path, then by the definition of the windability, it holds that
  \begin{align*}
    \prod_{v\in
    V}B_v(\sigma|_{\mathcal{E}(v)},\pi|_{\mathcal{E}(v)},M_v)
    =\;\prod_{v\in
    V}B_v((\sigma\oplus P)|_{\mathcal{E}(v)},(\pi\oplus
    P)|_{\mathcal{E}(v)},M_v),
  \end{align*}
  where we use $\sigma\oplus P$ to denote the configurations obtained
  from $\sigma$ by flipping the value on vertices in $P$.

  This finishes the proof by noting that
  $\tuple{\sigma\oplus P,\pi\oplus P}\in\Omega_2\times\Omega_2$ and
  the mapping $(\sigma,\pi)\to\tuple{\sigma\oplus P,\pi\oplus P}$ is
  injective.
\end{proof}

\begin{lemma}
  Let $\Gamma_0$ be the canonical paths from $\Omega_0$ to $\Omega$
  constructed above, then
  $\rho(\Gamma_0)\le \frac{n^3}{\mu_\Lambda(\Omega_0)}$.
\end{lemma}
\begin{proof}
  The congestion of $\Gamma_0$ is
  \[
    \rho(\Gamma_0)=\max_{(\sigma,\pi)}\frac{1}{\mu_\Lambda(\sigma)P(\sigma,\pi)}\sum_{\gamma\in\Gamma_0\mbox{\scriptsize~with
      }(\sigma,\pi)\in\gamma}\wt{\gamma}.
  \]
  By the definition of the Markov chain, it holds that
  $\mu_{\Lambda}(\sigma)P(\sigma,\pi)=\frac{1}{n^2}\min\tuple{\mu_\Lambda(\sigma),\mu_\Lambda(\pi)}$,
  thus
  \begin{align*}
    \rho(\Gamma_0)
    &\le \max_{\pi\in\Omega}\frac{n^2}{\mu_\Lambda(\pi)}\sum_{\gamma\in\Gamma_0\mbox{\scriptsize~with }\pi\in\gamma}\wt{\gamma}\\
    &\le \max_{\pi\in\Omega}\frac{n^2}
      {\mu_\Lambda(\pi)}
      \sum_{\substack{\sigma_1\in\Omega_0\\\sigma_2\in\Omega}}
    \sum_{\substack{\tuple{M_v\in\mathcal{M}_{z|_{\mathcal{E}(v)}}}_{v\in V}\\\mbox{\scriptsize~with }\pi\in\gamma_{\sigma_1,\sigma_2,M}}}\wt{\gamma_{\sigma_1,\sigma_2,M}}
    \quad\tuple{M\mbox{ is induced by }\tuple{M_v}_{v\in V}, z=\sigma_1\oplus\sigma_2}\\
    &=\max_{\pi\in\Omega}\frac{n^2}{w_\Lambda(\pi)(Z_0+Z_2)} \sum_{\substack{\sigma_1\in\Omega_0\\\sigma_2\in\Omega}}
    \sum_{\substack{\tuple{M_v\in\mathcal{M}_{z|_{\mathcal{E}(v)}}}_{v\in V}\\\mbox{\scriptsize~with }\pi\in\gamma_{\sigma_1,\sigma_2,M}}}\prod_{v\in V}B_v(\sigma_1|_{\mathcal{E}(v)},\sigma_2|_{\mathcal{E}(v)},M_v)\\
    &=\max_{\pi\in\Omega}\frac{n^2}{w_\Lambda(\pi)(Z_0+Z_2)} \sum_{\substack{\sigma_1\in\Omega_0\\\sigma_2\in\Omega}}
    \sum_{\substack{\tuple{M_v\in\mathcal{M}_{z|_{\mathcal{E}(v)}}}_{v\in V}\\\mbox{\scriptsize~with }\pi\in\gamma_{\sigma_1,\sigma_2,M}}}\prod_{v\in V}B_v(\pi|_{\mathcal{E}(v)},(\pi\oplus\sigma_1\oplus\sigma_2)|_{\mathcal{E}(v)},M_v)\\
    &\le \max_{\pi\in\Omega}\frac{n^2}{w_\Lambda(\pi)(Z_0+Z_2)} \sum_{\substack{\sigma_1\in\Omega_0}}\sum_{w\in\Omega_0\cup\Omega_2}\sum_{\substack{\tuple{M_v\in\mathcal{M}_{z|_{\mathcal{E}(v)}}}_{v\in V}\\\mbox{\scriptsize~with }\pi\in\gamma_{\sigma_1,\sigma_2,M}}}\prod_{v\in V}B_v(\pi|_{\mathcal{E}(v)},(\pi\oplus w)|_{\mathcal{E}(v)},M_v)
    \quad \tuple{w\defeq\sigma_1\oplus\sigma_2}\\
    &\le \max_{\pi\in\Omega}\frac{n^3}{w_\Lambda(\pi)(Z_0+Z_2)} 
      \cdot\sum_{w\in\Omega_0\cup\Omega_2}
      \prod_{v\in V}f_v(\pi|_{\mathcal{E}(v)})f_v((\pi\oplus w)|_{\mathcal{E}(v)})\\
    &\le n^2\cdot\frac{Z_0+Z_2+Z_4}{Z_0+Z_2}\\
    &\le \frac{n^3}{\mu_\Lambda(\Omega_0)},
  \end{align*}
  where the last inequality is due to Lemma \ref{lem:Z2bound}.
\end{proof}

\begin{lemma}
  Let $\Gamma$ be the canonical paths from $\Omega$ to $\Omega$
  constructed above, then
  $\rho(\Gamma)\le \frac{n^3}{\mu_\Lambda(\Omega_0)^2} $.
\end{lemma}
\begin{proof}
  The congestion of $\Gamma$ is
  \[
    \rho(\Gamma)=\max_{(\sigma,\pi)}\frac{\mu_\Lambda(\sigma)P(\sigma,\pi)}{\sum_{\gamma\in\Gamma\mbox{\scriptsize~with
        }(\sigma,\pi\in\gamma)}\wt{\gamma}}.
  \]
  By the definition of $\Gamma$, each $\gamma\in \Gamma$ is the
  concatenation of two paths in $\Gamma_0$. Denote $\mathbf{1}_A$ the
  indicator function of the event $A$, we have
  \begin{align*}
    \rho(\Gamma)
    &=\max_{(\sigma,\pi)}\frac{1}{\mu_\Lambda(\sigma)P(\sigma,\pi)}
      \sum_{\substack{x,z\in\Omega\\y\in\Omega_0}}\sum_{M_1\in\mathcal{M}_{x\oplus y}}\sum_{M_2\in\mathcal{M}_{z\oplus y}}
    \mathbf{1}_{\substack{(\sigma,\pi)\in \gamma_{x,y,M_1} \\ \lor(\sigma,\pi)\in \gamma_{z,y,M_2}}}\cdot \frac{\wt{\gamma_{x,y,M_1}}\wt{\gamma_{z,y,M_2}}}{\mu_\Lambda(y)\mu_\Lambda(\Omega_0)}\\
    &=\max_{(\sigma,\pi)}\frac{1}{\mu_\Lambda(\sigma)P(\sigma,\pi)}
      \sum_{\substack{x,z\in\Omega\\y\in\Omega_0}}\sum_{\substack{M_1\in\mathcal{M}_{x\oplus y}\\\mbox{\scriptsize~with }(\sigma,\pi)\in\gamma_{x,y,M_1}}}
    \sum_{M_2\in\mathcal{M}_{y\oplus z}}\frac{\wt{\gamma_{x,y,M_1}}\wt{\gamma_{z,y,M_2}}}{\mu_\Lambda(y)\mu_\Lambda(\Omega_0)}\\
    &=\max_{(\sigma,\pi)}\frac{1}{\mu_\Lambda(\sigma)P(\sigma,\pi)}
      \sum_{\substack{x,z\in\Omega\\y\in\Omega_0}}\sum_{\substack{M_1\in\mathcal{M}_{x\oplus y}\\\mbox{\scriptsize~with }(\sigma,\pi)\in\gamma_{x,y,M_1}}}
    \frac{\wt{\gamma_{x,y,M_1}}\mu_\Lambda(z)}{\mu_\Lambda(\Omega_0)}\\
    &=\max_{(\sigma,\pi)}\frac{1}{\mu_\Lambda(\sigma)P(\sigma,\pi)}
      \sum_{\substack{x\in\Omega\\y\in\Omega_0}}\sum_{\substack{M_1\in\mathcal{M}_{x\oplus y}\\\mbox{\scriptsize~with }(\sigma,\pi)\in\gamma_{x,y,M_1}}}
    \frac{\wt{\gamma_{x,y,M_1}}}{\mu_\Lambda(\Omega_0)}\\
    &=\frac{\rho(\Gamma_0)}{\mu_\Lambda(\Omega_0)}\\
    &\le\frac{n^3}{\mu_\Lambda(\Omega_0)^2}
  \end{align*}
\end{proof}



\end{document}